\documentclass[aps,pra,twocolumn,superscriptaddress]{revtex4-2}
\usepackage{amsmath,bm} 
\usepackage{graphicx}
\usepackage[colorlinks=true,urlcolor=blue,breaklinks=true]{hyperref}

\begin{document}
\title{Vortex dynamics in rotating dipolar supersolids across Josephson and self-trapping regimes}

\author{Aitor Ala\~na} 
\affiliation{Department of Physics, University of the Basque Country UPV/EHU, 48080 Bilbao, Spain} 
\affiliation{EHU Quantum Center, University of the Basque Country UPV/EHU, 48940 Leioa, Biscay, Spain}

\author{Michele Modugno} 
\affiliation{Department of Physics, University of the Basque Country UPV/EHU, 48080 Bilbao, Spain} 
\affiliation{EHU Quantum Center, University of the Basque Country UPV/EHU, 48940 Leioa, Biscay, Spain} 
\affiliation{IKERBASQUE, Basque Foundation for Science, 48009 Bilbao, Spain} 

\author{Pablo Capuzzi} 
\affiliation{\hbox{Universidad de Buenos Aires, Facultad de Ciencias
Exactas y Naturales, Departamento de Física, 1428 Buenos Aires,
Argentina}} \affiliation{CONICET - Universidad de Buenos Aires,
Instituto de Física de Buenos Aires (IFIBA), 1428 Buenos Aires,
Argentina}

\date{\today}
\begin{abstract}
We investigate vortex nucleation and transport in a rotating dipolar supersolid
arranged in a triangular droplet lattice, exploiting its description as an
array of weakly linked condensates. By considering both Josephson and
macroscopic self-trapping dynamics, we show that local phase differences
between droplets provide a compact and highly predictive framework to explore a
wide range of vortex behaviors. In particular, Josephson oscillations can be
devised to induce vortex nucleation and motion near the vertices of the
low-density hexagonal lattice (between droplets), while self-trapping dynamics
induce running phases that enable directed vortex transport, which may be
accompanied by vortex–antivortex pair creation and annihilation over finite
time scales. Comparison with simulations based on the extended Gross-Pitaevskii
equation demonstrates that a three-droplet description is essential to capture
vortex motion near hexagon vertices.
Together, Josephson and self-trapping dynamics provide a tunable protocol
to trigger and track vortex nucleation, transport, and vortex-antivortex pair
annihilation, revealing the microscopic topological mechanisms underlying phase
slips in rotating dipolar supersolids.
\end{abstract}
\maketitle

\section{Introduction}

The study of macroscopic quantum phenomena focuses on systems that exhibit
key features such as phase coherence and superfluidity,
phenomena particularly manifest in exotic states of matter, such as
supersolids. These systems uniquely feature the spontaneous breaking of two
continuous symmetries: gauge symmetry, resulting in superfluidity, and
translational symmetry, leading to a crystal-like density modulation. Recent
experimental breakthroughs have successfully realized dipolar supersolids
composed of self-sustained quantum droplets in ultracold atomic gases 
\cite{Ta19,bottcher19,ch19}, providing a versatile platform to explore the 
relation between the solid and superfluid properties
of this state. Understanding the response of these complex systems to rotation,
particularly concerning the emergence and dynamics of topological excitations,
such as quantized vortices, remains a central challenge. In this context,
it is worth noting that vortices in ultracold atomic gases can be directly
detected and tracked using established imaging techniques, and recent
experiments have demonstrated their observation both in atomic superfluids
\cite{HernandezRajkov2024} and in dipolar supersolids \cite{casotti2024}.

Supersolids formed by droplets have been shown to behave as weakly linked
condensates that can exhibit both Josephson and macroscopic quantum
self-trapping (ST) oscillations \cite{Biagioni2024,alana25,Donelli25,abad11a}.
Modeling these systems as weakly linked condensates, or a Josephson junction
array, is a powerful methodology for probing their transport properties
\cite{ragh99,albiez05,anan06,jezek13a}. The ST regime is characterized by a
population imbalance between different droplets that oscillates but never
crosses the equilibrium value and is accompanied by a monotonically increasing,
or ``running," phase difference.  This continuous increase in the phase of the
macroscopic wave function requires phase slips to occur repeatedly. Therefore,
the stability of the ST regime relies directly on these repeated phase slips.
Microscopically, these phase slips are associated with the dynamical creation,
motion, or annihilation of elementary topological excitations
\cite{abad11b,yaki2015,abad15,munoz2015,xhani2020}. 

In non-rotating systems, it has been shown that for the ST regime, vortices can
appear to produce these phase slips \cite{abad11b,abad15}. More precisely, the
passage of vortices along the junction has been shown to be related to the
local inversion of the velocity field in the surrounding region.  Such a
phenomenon happens in times very short with respect to the typical times
involved in the dynamics of the macroscopic variables, making the underlying
microscopic vortex dynamics difficult to detect and analyze. It is worthwhile
to recall that the phase slips can also occur through the appearance of a nodal
surface along the junction, depending on which effect is energetically favored.
Furthermore, in dipolar supersolids, the density structure profoundly
influences vortex behavior \cite{gal20, rocc20,casotti2024,poli23,aitor} often
leading to vortex pinning at low-density interstitial regions. 

The nucleation and dynamics of vortices have been vastly studied in different
trapping potentials in ultracold rotating systems
\cite{Castin1999,Madison2000,Aboshaeer2001,Modugno2003,williams10,ka11,je23}.
In a recent work \cite{rot20}, it has been shown that when the condensate can
be considered as constituted by well-defined, weakly linked condensates, with
each of them having an on-site almost axially symmetric density, the individual
velocity fields are homogeneous, and then, the phases acquire a linear
dependence on the coordinates. 
More precisely, the velocity field of the $i$-th condensate is given by
$\textbf{v}_i=\mathbf{\Omega} \times \textbf{r}_i$.  Recently, the
corresponding expression for the phases has proven to provide a powerful tool
for the determination of the position of vortices \cite{je23,aitor} in systems
of this type. In particular, for a square lattice, it has been shown that the
position and the dynamics of vortices along the low-density valley between two
neighboring sites can be described by a simple expression that considers only
the phases at those sites. By contrast, in the case of a triangular lattice, it
has been demonstrated by using a supersolid array of droplets that to describe
the stationary positions of vortices, the phases of three neighboring droplets
are needed, especially near the points that are equidistant from their centers
\cite{aitor}.

In this work, we investigate the vortex dynamics in a rotating dipolar
supersolid arranged in a triangular lattice of droplets, a geometry that allows
us to test both two- and three-droplet approximations during vortex motion. The
dynamics are generated by combining Josephson and self-trapping dynamics, which
enables a versatile exploration of the response of the supersolid. Our primary
aim is to demonstrate that the dynamics of vortices can be comprehensively
described using the local phase difference between neighboring droplets. In
particular, we also address the link between the macroscopic ST regime and the
microscopic topological excitations responsible for phase slips. We employ
full three-dimensional numerical simulations of the extended
Gross-Pitaevskii equation (eGPE) to investigate the dynamics and test our
theoretical model, leading to several key findings: we confirm the accuracy of
using local phase differences among droplets as a highly predictive tool for
modeling complex vortex trajectories within the supersolid lattice, validating
the necessity of the three-droplet approximation to accurately model motion
near the vertices of the hexagonal lattice formed by the low-density
interstitial regions between droplets.  Finally, we report the detailed
observation and characterization of vortex-antivortex pair creation and
subsequent annihilation occurring under specific ST oscillations. 

The paper is organized as follows. In Sec. \ref{sec:sys} we introduce the
specific system under study and the dynamical protocol used to nucleate and
investigate the vortex dynamics. Section \ref{sec:theo} presents the truncated
weakly linked condensates theory utilized to predict the position of vortices
along the different low-density paths, extending previous results to the
dynamical regime of self-sustained rotating supersolid. The numerical results
of the eGPE and their comparison with the presented theory are examined in
Sec.\ \ref{sec:num}, where the three-droplet approximation is validated for
vortex motion near hexagon vertices,  and the role of the stationary
configuration asymmetry in the model is discussed. Finally, Sec. \ref{sec:sum}
offers the summary and concluding remarks of our work.

\section{\label{sec:sys}System and dynamical protocol}

\begin{figure}[th]
 \includegraphics[width=\columnwidth]{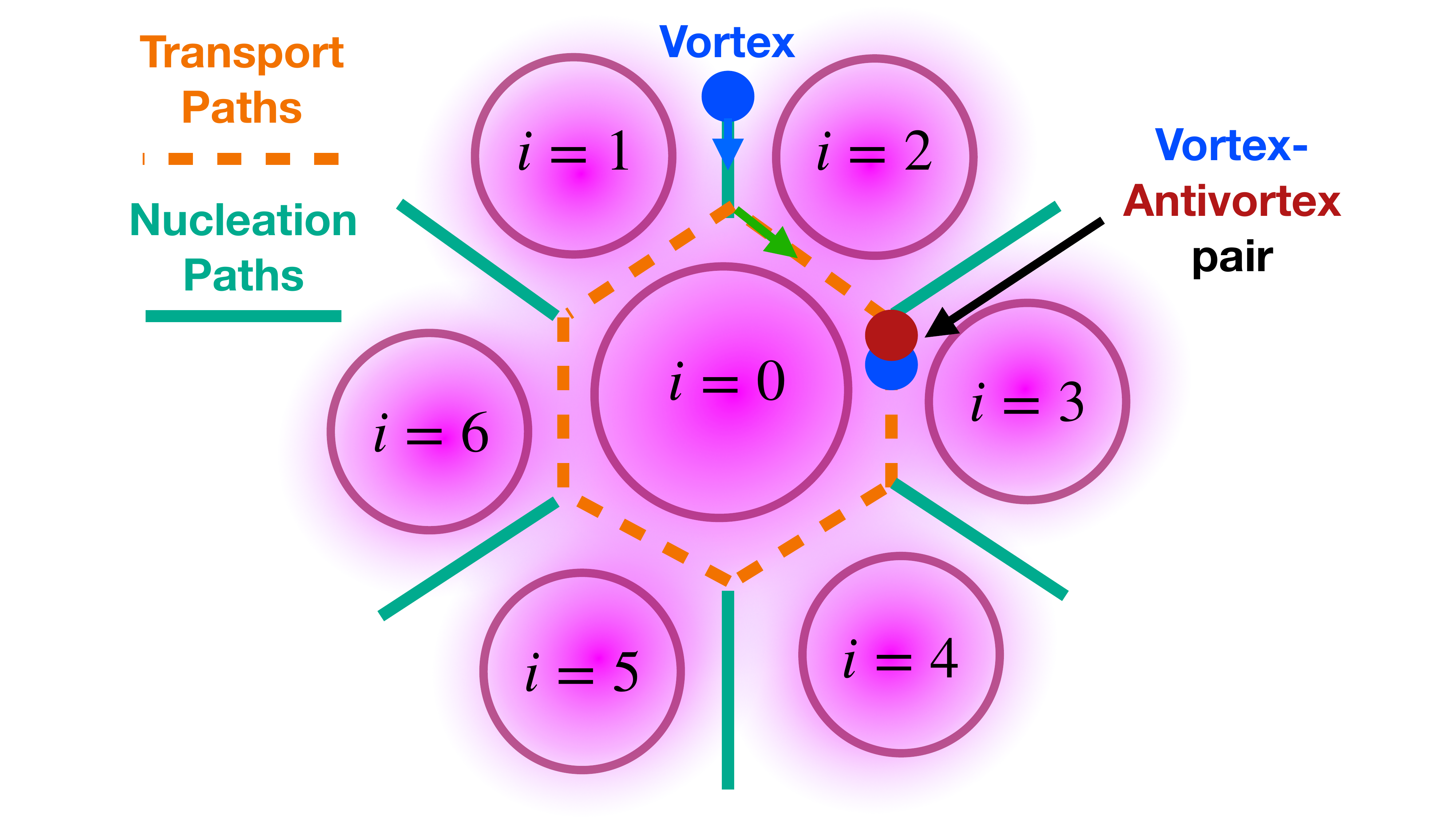}
 \caption{\label{fig:scheme} Schematic view of the system under study. 
 A supersolid with a central droplet and a ring of six smaller droplets,
arranged in a triangular lattice.  The low-density nucleation and transport
paths are highlighted in solid green and dashed orange, respectively. Vortices
(blue dots) can enter through nucleation, remaining there or moving to a
nucleation path. In the case of self-trapping, vortex-antivortex pairs can
appear, in which case the vortex and the antivortex (dark red dot) will move in
opposite directions until they collide with their opposite and annihilate.
 }
\end{figure}

We investigate the vortex dynamics of a dipolar Bose gas forming a
self-sustained supersolid arranged in a triangular lattice of seven droplets: a
central droplet ($i=0$) surrounded by a ring of six droplets ($i=1-6$), as
shown in Fig. \ref{fig:scheme}.  This configuration gives rise to Josephson
junctions connecting the droplets and is chosen for our analysis, as it
provides distinct low-density pathways for vortex entry and dynamics, enabling
a simultaneous analysis of mechanisms influenced by two droplets and those
influenced by three droplets.

To model the system, we consider an array of weakly linked condensates with a
total particle number $N=\sum_{i=0}^{6}N_{i}$.  We assume the system maintains
six-fold symmetry during the dynamics, implying the ring droplets share the
same population, $N_{i}(t)=N_{r}(t)$, and phase, $\phi_{i}(t)=\phi_{r}(t)$, for
$1\le i\le6$, measured at the center of the droplets.  The macroscopic dynamics
of the system is characterized by the normalized population imbalance $Z(t)$
and phase difference $\varphi(t)$ between the central and the ring droplets
\begin{align}
\label{eq:imbalance}Z(t)&=\frac{6N_{r}(t)-N_{0}(t)}{N}=1-2n_0(t)\\\
\label{eq:phasediff}\varphi(t)&=\phi_{0}(t)-\phi_{r}(t),
\end{align}
where $n_0=N_0/N$ is the fraction of atoms in the central droplet. We also
define $Z_e$ as the stationary population imbalance corresponding to the
ground-state configuration. These variables allow us to distinguish the
Josephson and ST regimes.

The role of the six-fold symmetry of the system is to enable a simpler
analytical description of the Josephson junction dynamics of the supersolid: it
allows the dynamics of a system of seven droplets to be captured by just $Z(t)$
and $\varphi(t)$. However, the predictive models for vortex position presented
throughout the paper do not rely on any specific geometrical configuration, as
they require only the populations and phases of the two or three droplets
closest to the vortex.

To provide a specific example, we consider a system composed of a gas of
$N\simeq1.1\times10^5$ $^{162}\mathrm{Dy}$ atoms at zero temperature and
confined by an external three-dimensional harmonic trap with frequencies
$\{\omega_{\perp},\omega_{z}\}=2\pi\times\{60,120\}$ Hz, as in previous
works \cite{alana25,gal20}. The dipolar scattering length is set to
$a_{dd}=130a_{0}$ and the $s$-wave scattering length is $a_{s}=92a_{0}$. This
system is capable of sustaining both Josephson oscillations and self-trapping
dynamics  \cite{Biagioni2024,alana25,Donelli25}.

The evolution of the system's wave function is governed by the eGPE, which
includes both the dipole-dipole interaction \cite{ronen2006} and quantum
fluctuations in the form of the Lee-Huang-Yang correction
\cite{fischer2006a,Li12,wachtler2016,schmitt2016}.

\subsection{Stirring protocol\label{sec:protocol}}

The rotation of the supersolid is  induced via a time-dependent confining
potential, as described in Ref. ~\cite{alana25}.  Specifically, in addition to
the axially symmetric harmonic trap described above, we superimpose an external
potential consisting of a set of Gaussian wells arranged with the same symmetry
as the ring of droplets. This potential, referred to as the \textit{egg-box}
potential, can be conveniently used to create a population imbalance in the
system and to set the system into rotation. It can be written as (see the
Supplemental Material of Ref. ~\cite{alana25})
\begin{equation}
\label{eq:egg}
\! V_{\text{egg}}(\bm{\rho},t)=V_0(t)e^{-2\rho^2\!/\sigma_{0}^2}{+V_{r}(t)\!\sum_{i=1}^{6}\!e^{-2|\bm{\rho}-\bm{\rho}_{0i}(t)|^2\!/\sigma_{0}^2}},
\end{equation}
with $\bm{\rho}$ representing the radial coordinate in the $xy$ plane,
$\bm{\rho}_{0i}$ the center-of-mass positions of the ring droplets,
$\sigma_{0}$ the widths of the Gaussians, and $V_0$ and $V_r$ the strengths of
the central and the ring wells, respectively. Both $V_0$ and $V_r$ are
negative, so $V_{\text{egg}}$ acts as an attractive set of wells that pin the
droplets.

\begin{figure}[t]
 \includegraphics[width= 0.9 \columnwidth]{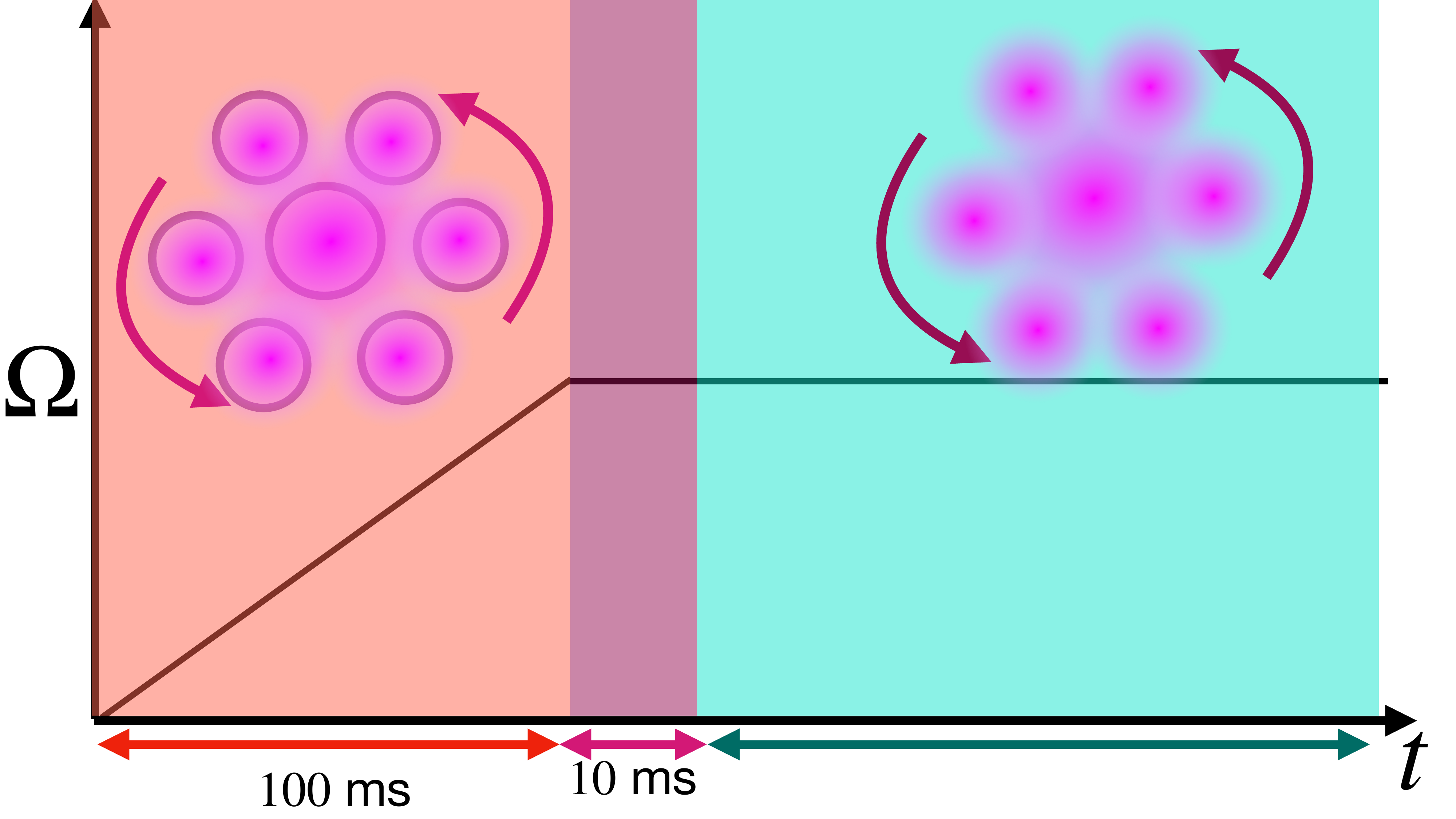}%
 \caption{\label{fig:protocol} Schematic depiction of the stirring protocol. 
 We start with a stationary self-sustained supersolid with a superimposed
egg-box potential.  This potential is then rotated at a linearly increasing
speed during 100 ms.  Following this, the potential is gradually turned off
over a period of 10 ms. Once the egg-box potential has been completely removed,
the supersolid rotates freely due to the angular momentum transferred by the
rotating potential.  }
\end{figure}

The rotating supersolid's evolution is engineered by precisely setting the
potential strengths, $V_0$ and $V_r$, which establish the relative population
imbalance during the preparation phase. The simulation sequence, shown in
Fig.~\ref{fig:protocol},  begins with a 100 ms ramp, linearly increasing the
angular velocity from zero to the final value, $\Omega$.  This is followed by a
10 ms relaxation ramp during which the egg-box potential is gradually switched
off to avoid sudden excitations. Once the potential is removed, the rotating
supersolid is allowed to evolve freely.  This free dynamics may result in
either a Josephson or self-trapping oscillation, which allows for the
exploration of various vortex dynamics.  It is worth noting that, near the
transition between the Josephson and self-trapping regimes, the dynamics
exhibit a mixed character: since the oscillation timescales become longer close
to the transition, successive cycles can alternate between the two types of
behavior, as the imbalance may shift slightly across the threshold from one
oscillation to the next. In what follows, we focus on cases that are clearly
within one regime or the other. To track the vortex dynamics resulting from
the numerical evolution of the eGPE, we utilize a plaquette method
\cite{Foster2010}.

\section{\label{sec:theo}Theoretical model for the vortex location}
In this section, we apply a truncated multimode model to describe the evolution
of the system. This type of model has been previously shown to be accurate in
describing the dynamics of conventional Bose-Einstein condensates in optical
lattices, capturing the main features of both Josephson and ST oscillations. 

The supersolid wave function $\psi(\mathbf{r},t)$ is approximated as the sum of
localized wave functions  $w_i(\mathbf{r},\Omega)$ centered at each droplet $i$
and normalized to unity. Hence, 
\begin{equation}
  \psi(\mathbf{r},t)=\sum_i|w_{i}(\mathbf{r},\Omega)| e^{i \frac{m}{\hbar} (\bm{\rho}-\bm{\rho}_{0i})\cdot(\mathbf{\Omega}\times\bm{\rho}_{0i})}\times \sqrt{N_i(t)}\, e^{i \phi_{i} (t)}, 
\label{eq:psii}
\end{equation}
with $\mathbf{\Omega}=\Omega \hat{z}$. 
The total phase associated with $w_i(\mathbf{r},\Omega)$ consists of a
geometrical term,
$m(\bm{\rho}-\bm{\rho}_{0i})\cdot(\mathbf{\Omega}\times\bm{\rho}_{0i})/\hbar$,
arising from the rotation about the supersolid center
\cite{rot20,alana24,rocc20} consistent with the partial rigid-rotation
response of a supersolid with finite superfluid fraction, and a local
contribution given by the phase $\phi_i$ relative to the central droplet
\cite{rot20,alana24}. The six-fold symmetry of this particular geometry allows
us to assume that the centers of all the droplets in the ring keep the same
phase \cite{alana24,alana25}. We found this description to be in agreement with
our numerical simulation of the eGPE. Specifically, the equivalence of the
ring-droplet phases is maintained by the symmetric nucleation protocol, and the
phases extracted from the six ring droplets remain equal to within numerical
precision throughout the dynamics, as illustrated by the eGPE phase maps in
Fig.~\ref{fig:pan_ST}. Moreover, the model presented here, and the
applicability of the results, can be easily extended to other lattice
geometries.

The position of a vortex in the $xy$ plane $\bm{\rho}_v=(X_v,Y_v)$ in the
supersolid is described by the fulfillment of a single condition: the total
wave function $\psi$ must vanish at such a position, i.e.,
$\psi(\bm{\rho}_v,z)=0$. Since at any given position the total wave function is
approximated as the sum of the localized wave functions of the surrounding
droplets, their contributions should cancel each other out.  In the geometry we
focus on, there are two types of paths for a vortex to travel through the
low-density valleys: the so-called nucleation paths from which vortices enter
the supersolid, depicted by solid green lines in Fig. \ref{fig:scheme},
and the transport paths around the central droplet, depicted by dashed
orange lines.

Throughout this work, all transport and nucleation paths are equivalent by
symmetry. We therefore choose as representatives the paths aligned along the
vertical $\hat{y}$ direction, corresponding to the nucleation (transport) path
between droplets 1 (0) and 2 (3). Each path is then parameterized by a single
coordinate $y$, defined such that the center of droplet $i=0$ lies at $y=0$.
This provides a consistent reference for comparing vortex positions along
different paths.

\subsection{\label{sec:nuclea} Vortex nucleation}

When a vortex enters the system, its position is affected only by two droplets,
and therefore it can be readily determined solely from their total phases [cf.
Eq.~\eqref{eq:psii}]. Using this truncated approximation, denoted as the
two-droplet approach, we can describe the entrance along any of the six paths
of the hexagonal geometry (cf. Fig.\ \ref{fig:scheme}). In particular, the
$Y_n$ vortex coordinate along the nucleation path between the droplets $1$
and $2$, namely, the path with $x = 0 $ and $y > d / \sqrt{3} $ where $d$ is
the interdroplet distance, reads \cite{aitor}
\begin{equation}
Y_n (t)= (  2l+1 ) \, \frac{ \pi
 \hbar}{ m d \Omega} \,,
\label{vortn}
\end{equation}
with $l$ an integer number, labeling successive vortices along the same
nucleation path. Specifically, $l=0$ corresponds to the first-entering vortex,
i.e., the one closest to the center. Equation  (\ref{vortn}) can also be read
as a threshold condition on $\Omega$: the $l$-th vortex along a nucleation path
fits inside the cloud (i.e., at a position smaller than the outer radius
$R_\perp$) only when $\Omega > \Omega_l \equiv (2l + 1)\pi\hbar/(m d R_\perp)$.
For our parameters, the first vortex enters at $\Omega_0/2\pi \simeq 10$ Hz,
and the second at $\Omega_1/2\pi \simeq 27$ Hz (see also Ref. \cite{aitor}).
We note that because of the six-fold symmetry, the local phase difference
$\phi_1-\phi_2$ of the two droplets, which contributes to the total phase
difference used for location determination, vanishes. In the time-dependent
process, such nucleation dynamics hold if the system is driven adiabatically by
smoothly increasing the rotation frequency $\Omega(t)$, such that the atoms
remain well localized in droplets. When the egg-box potential is removed, the
vortex remains near such a value if the initial $d$ and $n_0$ values are close
to the equilibrium ones.  However, as the droplet populations evolve, their
sizes and distances between droplets change over time. As a result, and also
due to the small fluctuations of the free rotation, $Y_n$ is expected to
oscillate slightly around the position predicted by Eq. (\ref{vortn}).

Close to the intersection of a nucleation and a transport path, namely a vertex
of the hexagon depicted in Fig.\ \ref{fig:scheme}, the vortex is influenced by
three droplets: the central one and two ring droplets. For stationary
configurations, as shown in Ref. \cite{alana24}, the local phase differences
among the central droplet and the lateral ones are zero at low rotation
frequencies, reflecting the vanishing inter-droplet current among them.
This allows the vortex position $Y_v$ to be determined with good accuracy in a
semi-analytic manner using a three-droplet approximation that assumes a
Gaussian shape for the localized wave functions.  However, in a non-stationary
setting, the three-droplet approach must incorporate the variation of $\varphi$
in time alongside the position of the vortex in the $xy$ plane. Therefore, its
solution is obtained numerically by finding the zeroes of the superposition of
three Gaussian-shaped wave functions, using the actual time-dependent phases
and populations extracted directly from the eGPE results.

\subsection{ Vortex dynamics around the central droplet}

Along the lateral transport paths, the main contribution to the dynamics is
given by the time dependence of the phase difference between the central
droplet and the surrounding ones. In particular, for the vortex along the path
between the droplets $0$ and $3$, with $x \simeq d/2$ and $-d \sqrt{3}/6 <y<d
\sqrt{3}/6 $, the two-droplet approximation for the vortex position $Y_t$
reads \cite{je23,aitor},
\begin{equation}
Y_t (t)= \left( \frac{\varphi(t)}{\pi} + \, 2l+1 \right) \, \frac{ \pi
 \hbar}{ m d \Omega} \,,
\label{vortd}
\end{equation}
with $l$ an integer number, and here, the phase difference is given by
$\varphi(t)=\phi_0(t)-\phi_3(t) $. Although the two-droplet approximation is
less accurate for describing the dynamics when the vortex approaches the
vertex, one can use it to estimate a lower bound for $ | \varphi(t)| $ for a
vortex to appear in the lateral path, by assuming that such a vortex is at the
vertex. This yields the condition
\begin{equation}
 \left| \frac{\varphi}{\pi}\right|  >  1 - \, \frac{ m d^2 \Omega}{\sqrt{3} \hbar 2 \pi} \,.
\label{convarphi}
\end{equation}
In our system setup, such an estimate gives a bound $\left|
{\varphi}/{\pi}\right| = 0.45$ ($0.65$) for $\Omega/2 \pi= 30$ Hz ($\Omega/2
\pi= 20$ Hz). Hence, for Josephson oscillations, if the phase-difference
turning points are below such a value, no dynamics along the lateral paths
should be expected. On the other hand, for ST oscillations, a vortex dynamics
is always present since the value $ \left|{\varphi}\right| = \pi$ is always
reached. In the above selected path, this position corresponds to vortex
coordinates $X_t=d/2$ and $ Y_t = 0 $, where the density profile possesses a
saddle point. 
\begin{figure*}[t!]
\includegraphics[width=\textwidth]{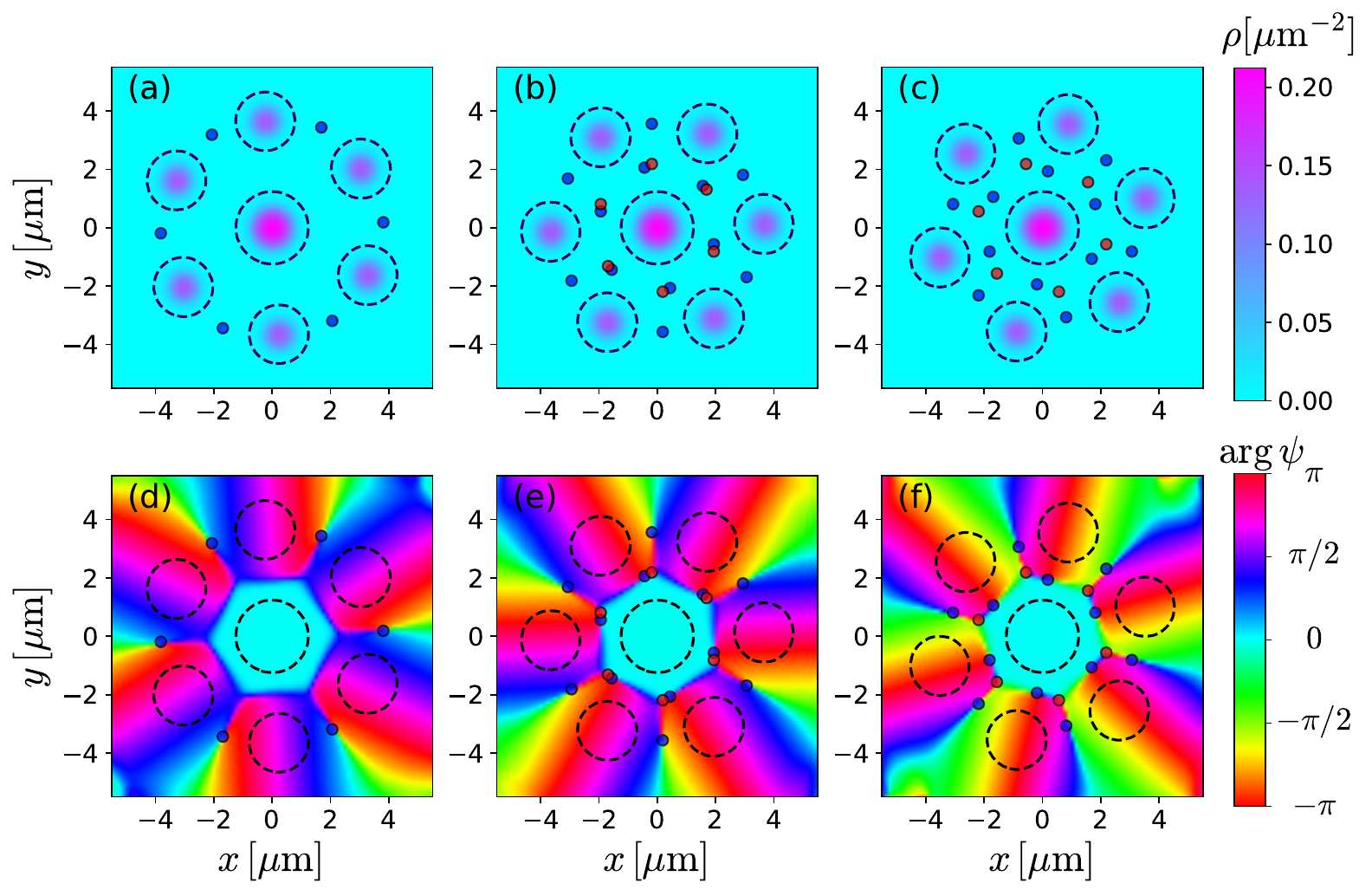}
\caption{\label{fig:pan_ST} Column densities $\rho(x,y)$
  (normalized to unity, top row, panels (a)-(c)) and wave
  function phases at $z=0$, $\arg \psi(x,y,z=0)$ (bottom row,
  panels (d)-(f)) of a rotating supersolid following the
  rotation ramp-up and a self-trapping dynamics. The dashed lines
  mark the density contour at $\rho=\rho_{\text{max}}/100$ 
    chosen to visualize the droplets cores, while the blue and
  red points mark vortices and antivortices, respectively.
  The actual  saddle-point column density between neighboring
  droplets is $\rho/\rho_{\text{max}}\simeq 10^{-5}$ and the
  superfluid fraction is approximately 7\%. The left, middle,
  and right columns correspond to times $t=70, 76, 84$ ms,
  respectively. }
\end{figure*}

As previously demonstrated for conventional BECs in external square lattices
\cite{capuzzi25}, the periods of the vortex orbits during a Josephson
oscillation are related to those of the population imbalance and relative
phases within a two-mode model. In the present supersolid configuration, 
the nonzero equilibrium population imbalance $Z_e \approx 0.59$ requires
the use of the asymmetric two-mode model of Refs. ~\cite{cat14,alana25}. Using
this model, the velocity of the vortex in the self-trapping regime can then be
approximated by
\begin{equation}
\dot{Y}_t (t)= \dot{\varphi}(t) \frac{ \hbar }{ m d \Omega} \simeq 
\frac{ 2 \pi \hbar }{ m d T \Omega} \, \mathop{\mathrm{sign}} \left( \Delta N_0 (t)\right),
\label{vortp}
\end{equation}
where $T$ is the corresponding ST oscillation period and $\Delta N_0(t)= N (
{n_0}_e - n_0(t) )$ is the difference between the number of particles in the
central droplet at the stationary state (${n_0}_e$) and its dynamical value
($n_0$).  Since in the ST regime the sign of $\Delta N_0(t)$ is constant, if
the number of particles in the central droplet is smaller (larger) than its
equilibrium value, vortices will move at roughly constant speed in the
counterclockwise (clockwise) direction around the central droplet without
oscillations. The time $T^*$ that a vortex spends to complete the passage along
a single lateral path can be thus estimated as 
\begin{equation}
T^* \simeq  \frac{ T d^2 \, \, (\Omega / 2 \pi) }{   \sqrt{3} \, (\hbar /m)}, 
\label{tp}
\end{equation}
incorporating dynamical as well as geometrical effects through $\Omega$, $d$, and $T$.

\section{\label{sec:num}Numerical simulations}
The numerical simulations are performed by using the full three-dimensional
eGPE for dipolar supersolids, as discussed in Refs.
\cite{alana25,alana22,Politi22}.  In this section, we present and
quantitatively compare the results for the vortex dynamics in the supersolid to
the theoretical model of the previous section. Figure \ref{fig:pan_ST} shows an
example of such a dynamics during the ramp-up of the rotation in the laboratory
frame, starting with initial conditions in the ST regime.  As seen, vortices
nucleate from the outside and enter the central droplet along the nucleation
trajectory described in Sec. \ref{sec:nuclea}, in accordance with the
stationary case studied in Ref. \cite{alana24}.  Furthermore, in the ST regime,
vortex–antivortex pairs may be nucleated to facilitate the transfer of the
dynamics from the nucleation paths to the lateral transport paths around the
central droplet.

Hereafter, the analysis is performed in the corotating frame of the supersolid,
as defined by the evolution of the droplet centers.  We shall demonstrate that
indeed vortices travel through the described paths and that while the vortex is
far from a hexagon vertex, the two-droplet formula provides an accurate
description. Then, for positions near the vertices, the three-droplet
correction is required to describe the dynamics.

\subsection{Josephson oscillations}

To obtain the Josephson dynamics, we linearly ramp the frequency starting from
a state close to the stationary configuration, with $Z=Z_e\simeq 0.6$ and
$\phi_i=0$. Hence, during this slow vortex nucleation process, the imbalance
$Z(t)$ evolves, determined by the presence of the egg-box potential.  When
the final frequency is reached, and the potential is removed, the imbalance
oscillates around an almost constant value $Z_e$.

\begin{figure}[h!]
 \includegraphics[width=\columnwidth]{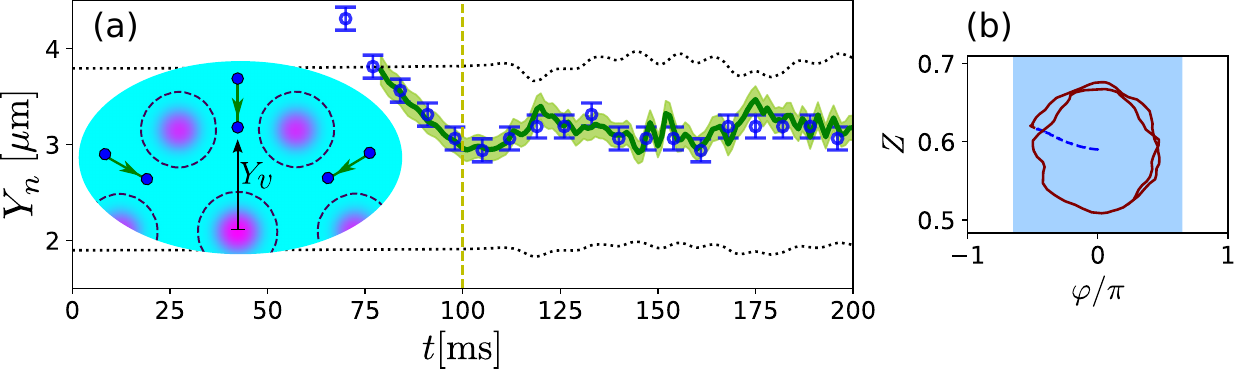}%
 \caption{\label{vort20jose}(a) Vortex nucleation $Y_n$ as a function of
time, following the stirring procedure in Fig.~\ref{fig:protocol}, for $\Omega/
2 \pi =20$ Hz and $Z_i\simeq 0.59$.  The eGPE results are indicated with blue
circles, and the two-droplet approximation with a solid green line.  The error
bars correspond to the spatial discretization in the numerical simulation,
while the colored band around the two-droplet approximation marks a 5\%
uncertainty stemming from the determination of $d$ and $\Omega$ from the
spatial discretization and averaging over the six ring droplets. The dotted
lines mark the position of the hexagon vertices. The inset illustrates
schematically the geometry of the droplets and the involved path $Y_n$.
(b) Evolution of the imbalance $Z$ and phase difference $\varphi$
extracted from the eGPE simulation. Dashed lines denote the ramp-up period, and
solid lines represent the free evolution. The colored region marks the
condition in Eq.~(\ref{convarphi}).}
\end{figure}

We first analyze a nucleation-only case in which the vortex dynamics along
lateral paths are absent. This  is obtained with an initial imbalance
$Z_i\simeq0.59$, $d\simeq 3.3\,\mu$m, and a final frequency $\Omega/2\pi=20$
Hz. The vortex nucleation shown in Fig. \ref{vort20jose} is well described by
the two-droplet approximation. In panel (b) of the figure, we can observe
the evolution of the imbalance and phase difference, showing that the lower
bound for the phase for obtaining a vortex dynamics around the vertices is not
reached, since its largest value is about $\varphi/ \pi \simeq 0.5$. Therefore,
no transport dynamics around the central droplet should be expected. It is also
confirmed that the oscillation is roughly around a constant $Z_e$.

\begin{figure}[h!]
\includegraphics[width=\columnwidth]{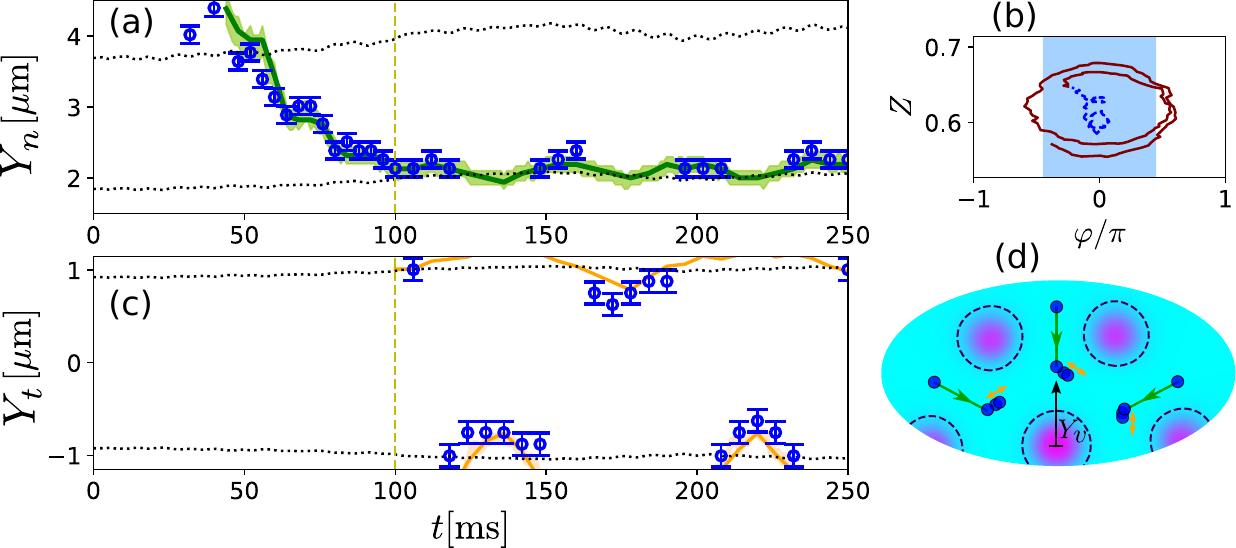}%
 \caption{\label{fig:vor30} (a) Vortex nucleation $Y_n$  and (c) vortex
dynamics along the transport path $Y_t$ as functions of time, for $\Omega/ 2
\pi =30$ Hz and during a Josephson oscillation. Results of the eGPE
simulations are indicated with blue circles, and the three-droplet
approximation with green (panel (a)) and orange (panel (c)) solid
lines.  The errors and right panels (b) and (d) are depicted as for Fig.\
\ref{vort20jose}, with panel (d) offering a schematic depiction of
vortex trajectories. }
\end{figure}

To induce dynamics along the lateral path towards a vertex, the final frequency
is increased to $\Omega/ 2 \pi =30$ Hz, such that the estimated lower bound of
$\varphi$ diminishes. In this case, as $Z$ and $\varphi$ are within the
Josephson regime, $\varphi$ has turning points, and then the nucleated vortex
that has already reached the vertex oscillates around it. This dynamics is
depicted in Fig. \ref{fig:vor30}. As seen, the dynamics after the removal of
the egg-box potential develop around the vertices. In panel (a) we observe
the entrance of vortices for $t<100$ ms towards the vertex, and its subsequent
oscillation both around the lateral and nucleation paths, as evinced in
panels (a) and (c).  It is worth mentioning that  panel  (c) also
captures the oscillation of another vortex around a second hexagon vertex. Due
to their motions around such vertices, the two-droplet approximation becomes
inaccurate, and the three-droplet approach becomes necessary to describe such
oscillations (green and orange solid lines in the figure).

\begin{figure}[!ht]
\centering
 \includegraphics[width= \columnwidth]{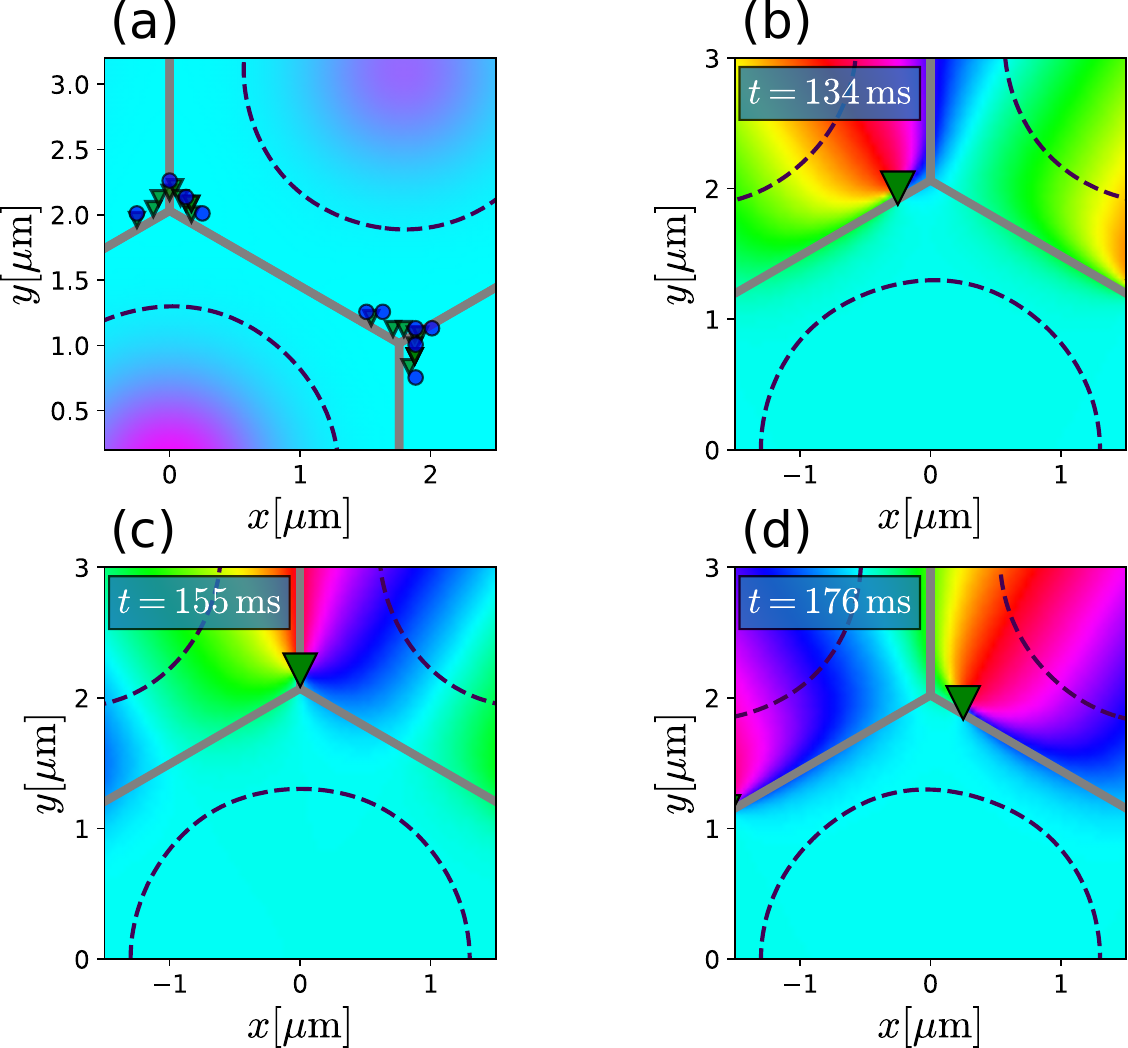}%
 \caption{\label{traj} Vortex oscillations in the $xy$ plane for the simulation
of Fig.\ \ref{fig:vor30}.  Panel (a) depicts the vortex positions at
different times during the dynamics according to the eGPE (blue dots) and
three-droplet approximation (green triangles) around two vertices, with the
background colors depicting the corresponding densities at $t=110$ ms.
Panels (b)-(d) show the position of the vortex that has been nucleated
along the $y>0$ axis up to the vertex, at selected observation times
$t=134, 155, 176$ ms, calculated using the three-droplet approximation,
with the background colors corresponding to the wave function phase.}
\end{figure}
The dynamics under study involve motion near the vertices, specifically along
two distinct paths corresponding to nucleation and transport. Accordingly, for
a more detailed analysis, panel (a) of Fig. \ref{traj} compares the
numerical simulation results with the predictions of the three-droplet
approximation, presented jointly in the $xy$ plane over a finite time interval.
The blue dots correspond to eGPE simulations, while the green triangles
represent the three-droplet approximation. The panels (b)-(d) in
Fig.~\ref{traj} show snapshots of the vortex positions at selected times, as
indicated in the figure. As seen in Fig. \ref{traj}, the three-droplet
approximation correctly describes the trajectory and the oscillation around a
vertex among droplets with $i=0, 1$, and $2$. 

\subsection{Self-trapping dynamics and creation of vortex-antivortex pairs} 

\begin{figure}[h!]
\includegraphics[width= \columnwidth]{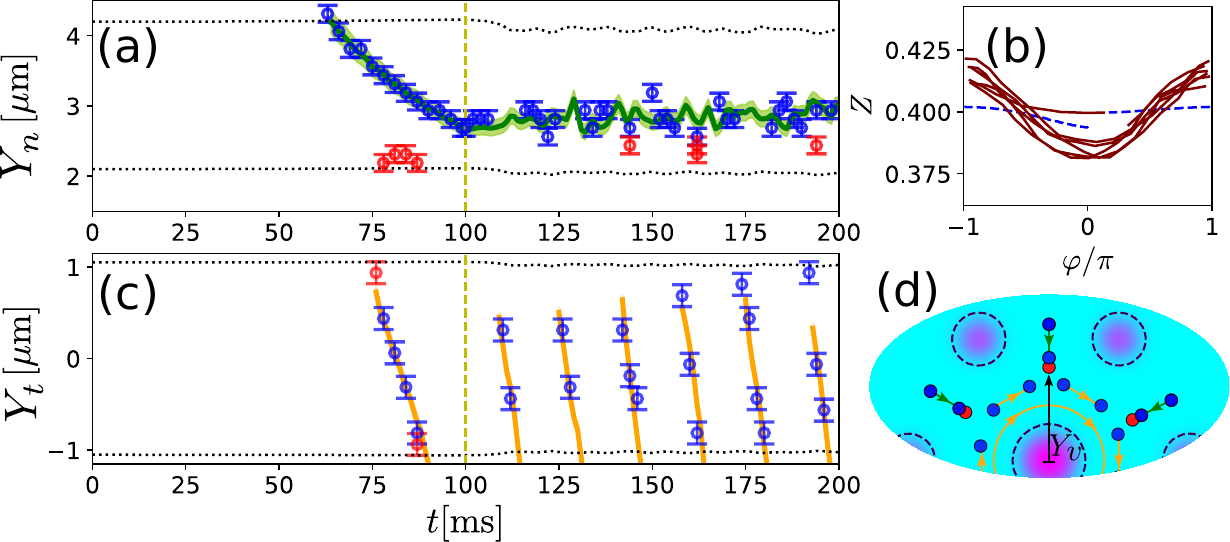}%
  \caption{\label{vort20abajo}
 (a) Vortex nucleation  $Y_n$ and (c) vortex dynamics along the lateral
transport path $Y_t$ as functions of time, for final $\Omega/ 2 \pi =20$ Hz.
The blue (red) circles correspond to vortices (antivortices) in the GP
simulation. The green solid line in  panel (a) and the orange solid line
in panel (c) are the predicted values using the two-droplet approximation.
Panel (b) depicts the imbalance $Z$ and phase difference $\varphi$ during
the ramp-up of the frequency (dashed lines), and the free evolution (solid
lines). Panel (d) shows a schematic depiction of the trajectories of
vortices and the position where vortex-antivortex nodes are created (although
not depicted, antivortices flow in the opposite direction to vortices, counter
clockwise).}
\end{figure}

To induce the self-trapping dynamics, we proceed in the same manner as for the
Josephson oscillations, but starting from a configuration farther from the
stationary state that the system would have without the egg-box potential.  In
particular, we first choose an initial imbalance of $Z_i=0.39$ and a final
rotation frequency $\Omega/2\pi=20$ Hz. This imbalance corresponds to the
central droplet being more populated than in the stationary configuration.
Figure \ref{vort20abajo} shows such a dynamics. As seen in panel (a), the
vortex initially follows the predicted evolution along the nucleation path
during the ramp-up phase (up to $t=100$ ms), where the interdroplet distance is
fixed. Subsequently, the positions of the vortices in the nucleation path
become  scattered with respect to the horizontal line due to the droplets'
radial movements.

\begin{figure}[!ht]
 \includegraphics[width= \columnwidth]{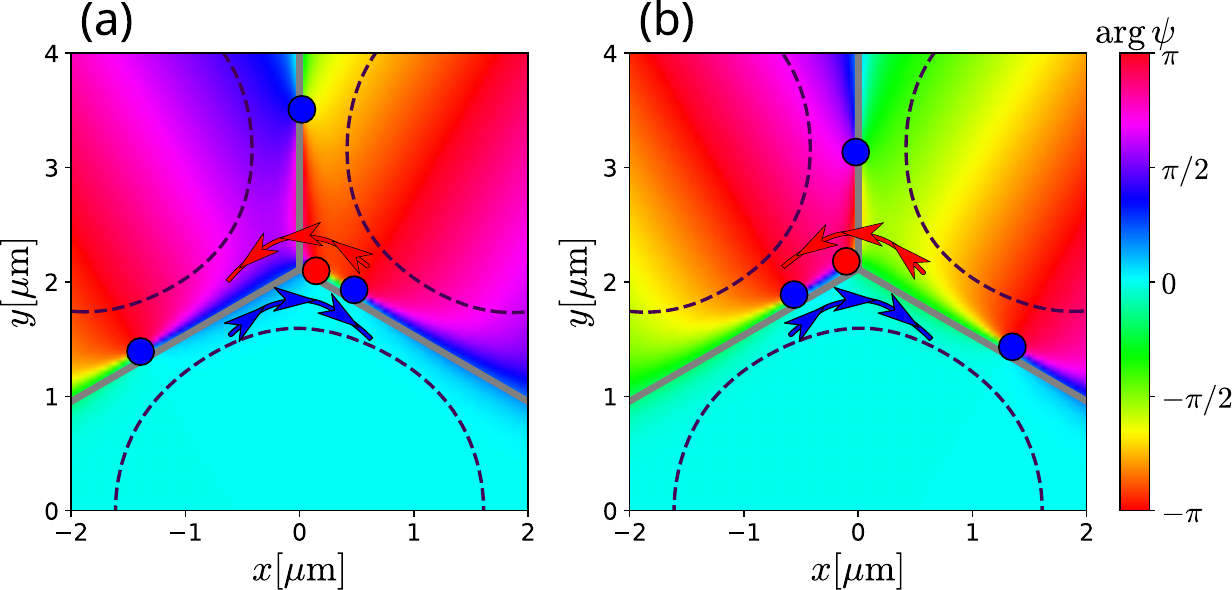}%
 \caption{\label{anti77} Position of the vortices (blue circles) and
antivortices (red circles) using the three-droplet approximation. Panel
(a) shows a vortex-antivortex at $t =77 $ms near its creation, and panel
(b) depicts the same antivortex some milliseconds after ($t = 85 $ms)
before they annihilate each other. The arrows indicate the direction in which
the vortices and antivortex move. The results correspond to the same simulation
of Fig. \ref{vort20abajo}. }
\end{figure}

Notably, for these initial conditions vortex–antivortex pairs are 
nucleated from the beginning of the ST dynamics. In the ST regime, 
the relative phase $\varphi$ between the central and ring droplets 
winds continuously by $2\pi$ over each oscillation period, and the 
corresponding $\pi$ phase slip must be physically realized along the 
lateral transport path that connects two neighboring vertices. When 
the geometry of the nucleation is such that the initially nucleated 
vortex cannot accommodate this slip on its own, typically because it 
remains too far from the vertex during the relevant time window, the 
slip is provided instead by the local creation of a vortex–antivortex 
pair at the vertex itself. 
A similar mechanism, in which vortex–antivortex pairs nucleate at 
low-density regions to accommodate a $\pi$ phase slip, has been 
previously reported in related Josephson 
settings~\cite{abad11b,yaki2015}. In our setup, the rotation of the 
system slows the subsequent motion of these pairs, allowing us to 
observe them with a lifetime on the order of milliseconds (see 
Fig.~\ref{vort20abajo}). Each resulting vortex and antivortex travels 
in opposite directions until the vortex meets and annihilates with 
the antivortex of a neighboring pair. 
Near the lower vertex, the first antivortex (red circles in Fig.
\ref{vort20abajo}) appears already during the nucleation process; the
resulting vortex then proceeds along the lateral path, as seen in panel (c).

The passage of such vortices occurs in a time $T^*$ that is very short, and can
be estimated using Eq.~(\ref{tp}), ($T \simeq 17$ ms, $\hbar/m=0.39 \mu{\rm
m^2/ms}$, $d=3.5 \mu$m), yielding $T^* \simeq 6.8$ ms. Also, a clockwise vortex
motion, as previously discussed, is a signal that the central droplet has more
particles than in the stationary state.  Figure \ref{anti77} clearly
illustrates this process during the nucleation phase:  panel (a) shows a
vortex-antivortex pair near its creation ($t=77$ ms), and  panel (b)
depicts the same antivortex milliseconds later ($t=85$ ms) before annihilation.
The arrows indicate their motion.

The appearance of such antivortices is not captured by the two-droplet
approximation, whereas, as seen in Fig. \ref{anti77}, it is correctly described
by the three-droplet approximation. As the positions of vortices or
antivortices are predicted by a zero of the wave function, they appear in the
graph as a phase difference of $ \pi $.  One can observe the little
deformations of the phases that provide the three-droplet approximation for
such a phase difference at the position of the antivortex too.

\begin{figure}[h!]
 \includegraphics[width= \columnwidth]{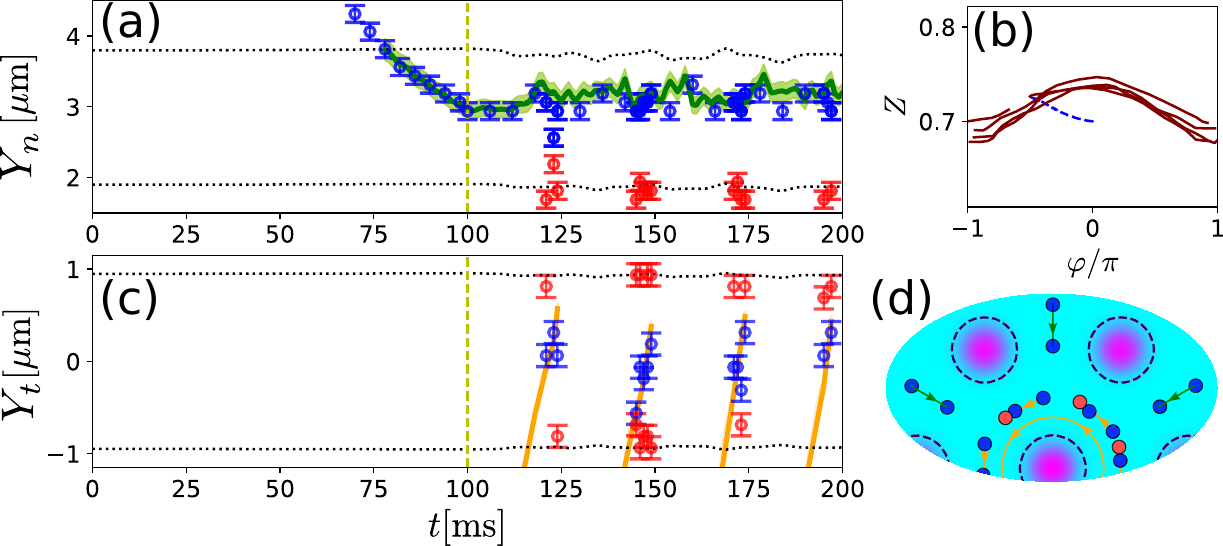}%
 \caption{\label{fig:vort20arriba}
(a) Vortex nucleation $Y_n$  and (c) vortex dynamics along the lateral
transport path $Y_t$  as functions of time, for initial imbalance $Z_i\simeq
0.7$ and final $\Omega/ 2 \pi =20$ Hz. The blue (red) circles correspond to
vortices (antivortices) in the eGPE simulation. The green solid line in panel
(a) and the orange solid line in  panel (c) are the predicted values
using the two-droplet approximation. Panel (b) depicts the imbalance $Z$
and phase difference $\varphi$ during the ramp-up of the frequency (dashed
lines), and the free evolution (solid lines). 
Panel (d) is a
schematic depiction of vortex trajectories, constructed as in Fig.
\ref{vort20abajo}.
 } 
\end{figure}

The creation and annihilation of vortex-antivortex pairs can be observed more
clearly in the simulation depicted in Fig.\ \ref{fig:vort20arriba}, where the
central droplet has a lower number of particles with respect to that of the
equilibrium one, and then vortices move counterclockwise. Such a simulation was
prepared with an initial imbalance $Z_i\simeq 0.7>Z_e$. In this case,
vortex-antivortex pairs emerge only during the free ST dynamics, and their
observation is facilitated by the larger time of passage $T^*$, resulting from
the longer ST oscillation period \cite{alana25}.

\begin{figure}[ht!]
 \includegraphics[width= \columnwidth]{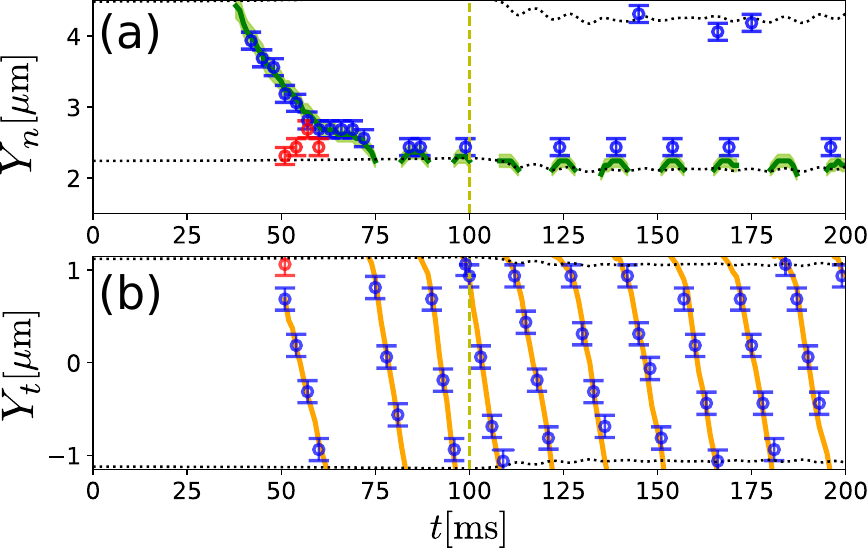}%
 \caption{\label{fig:vort30abajo}
 (a) Vortex nucleation $Y_n$ and (b) vortex dynamics along the lateral
transport path $Y_t$ as functions of time, for initial imbalance $Z_i\simeq
0.44$ and final $\Omega/ 2 \pi =30$ Hz. The blue (red) circles correspond to
vortices (antivortices) in the eGPE simulation. The green solid line in  panel
(a) and the orange solid line in  panel (b) are the predicted values
using the three-droplet approximation. }
\end{figure}

Finally, we consider the self-trapping dynamics induced after a nucleated
vortex reaches a vertex during the stirring procedure. This regime is achieved
for a final frequency $\Omega/2\pi= 30$ Hz  and an initial imbalance
$Z_i=0.44$. The results are displayed in Fig.\ \ref{fig:vort30abajo}.  As
observed, during the intermediate stage of the stirring process, when the
vortex first approaches a vertex, the transfer to a lateral path is accompanied
by the nucleation of a vortex–antivortex pair, analogously to what was
discussed for $\Omega/2\pi=20$ Hz; subsequently, during the free evolution, the
vortex circulates along the transport paths without further pair formation and
with a roughly constant speed, as indicated by the fixed slopes in panel
(b), consistent with the three-droplet description.

\section{\label{sec:sum}Summary and concluding remarks}

We investigated the vortex nucleation and dynamics in a rotating dipolar
supersolid confined within a harmonic trap, where the self-sustained droplets
form a triangular lattice exhibiting hexagonal symmetry.  By modeling the
system as an array of weakly linked condensates, we were able to predict the
vortex dynamics based on the knowledge of the phase differences between
neighboring droplets. As the rotation frequency was slowly increased, we
confirmed that vortices entered the system along the radial low-density paths,
in agreement with the theoretical predictions. Along the lateral paths
surrounding the central droplet, a more complex vortex dynamics was observed,
which depends sensitively on the initial conditions. 

We found that a minimum phase difference is required to induce such a lateral
dynamics, and we determined a threshold value that decreases as the frequency
increases. While a simpler two-droplet approximation is useful to describe
vortex motion in the low-density valleys when the vortex is far from a hexagon
vertex, we demonstrated that a three-droplet approach is essential for
accurately describing the system dynamics close to such vertices.

The observed dynamics included both oscillations around the hexagon vertices
and the passage of vortices between neighboring vertices. We found that the
creation of vortex-antivortex pairs can significantly facilitate this passage
process, particularly when nucleated vortices remain spatially separated from
the hexagon vertices. System rotation was shown to slow down the pair
trajectories, making these pairs observable in our simulations for several
milliseconds, long enough to track their creation and later annihilation along
adjacent paths. These observations further highlight the importance of the
multi-droplet modeling for reconstructing vortex motion near vertices and to
resolve the vortex–antivortex pair process that may accompany self-trapping
dynamics in this dipolar supersolid system. Additionally, the observed passage
of single vortices exhibited a remarkable quantitative feature: their velocity
was found to be inversely proportional to the rotation frequency, with its sign
dictated by the particle number difference between the central droplet and its
equilibrium value. 

In conclusion, this work provides a strong validation for the use of weakly
linked condensate models in supersolid systems and offers quantitative measures
for vortex velocity and nucleation thresholds. Crucially, we have shown that
Josephson and ST dynamics, generated by a controlled population imbalance,
provide a tunable protocol to trigger vortex and antivortex events in a
predictable way. This enables the systematic characterization of vortex
trajectories and the identification of the microscopic mechanisms underlying
phase slips, thereby enhancing our understanding of transport and topological
excitations in rotating dipolar supersolids.  These results remain directly
relevant to current experiments, where vortices can be detected and tracked
through established imaging techniques \cite{HernandezRajkov2024,casotti2024}.

\section{Acknowledgments}
We gratefully acknowledge Dora Jezek for her contributions to this work.  This
work was supported by Grant PID2021-126273NB-I00 funded by
MCIN/AEI/10.13039/501100011033 and by “ERDF A way of making Europe”, by the
Basque Government through Grant No. IT1470-22, and by the European Research
Council through the Advanced Grant “Supersolids” (No. 101055319).  P.C.
acknowledges support from CONICET and Universidad de Buenos Aires, through
grants PIP 11220210100821CO and UBACyT 20020220100069BA, respectively.


%

\end{document}